\documentclass[preprint2]{aastex}
\usepackage{graphicx}
\usepackage[colorlinks=true,citecolor=blue]{hyperref}
\bibpunct{(}{)}{;}{a}{}{,}
\usepackage[switch,pagewise]{lineno}
\usepackage{subfigure}
\usepackage{overpic}
\usepackage{multirow}
\usepackage{ulem}
\usepackage{color}
\usepackage{amsmath}
\usepackage{txfonts}
\usepackage{cases}
\usepackage{natbib}

\shorttitle{Effect of sausage oscillations on radio zebra-patterns}
\shortauthors{Yu, Nakariakov \& Yan}

\begin{document}

\title{Effect of a sausage oscillation on radio zebra-pattern structures in a solar flare}
\author{Sijie Yu$^1$, V.~M. Nakariakov$^{2,3,4}$, Yihua Yan$^1$}
\affil{$^1$Key Laboratory of Solar Activity, National Astronomical Observatories Chinese Academy of Sciences, Beijing 100012, China\\
	$^2$Centre for Fusion, Space and Astrophysics, Physics Department, University of Warwick, Coventry, CV4 7AL, UK\\
	$^3$School of Space Research, Kyung Hee University, Yongin, 446-701, Gyeonggi, Korea\\
	$^4$Central Astronomical Observatory at Pulkovo of RAS, 196140 St Petersburg, Russia}
\email{sjyu@nao.cas.cn}

\begin{abstract}
Sausage modes that are axisymmetric fast magnetoacoustic oscillations of solar coronal loops are characterized by variation of the plasma density and magnetic field, and hence cause time variations of the electron plasma frequency and cyclotron frequency. The latter parameters determine the condition for the double plasma resonance (DPR), {which is} responsible for the appearance of zebra-pattern (ZP) structures in time spectra of solar type IV radio bursts. We perform numerical simulations of standing and propagating sausage oscillations in a coronal loop modeled as a straight, field-aligned plasma slab, and determine the time variation of the DPR layer locations. Instant values of the plasma density and magnetic field at the DPR layers allowed us to construct skeletons of the time variation of ZP stripes in radio spectra. In the presence of a sausage oscillation, the ZP structures are shown to have characteristic wiggles with the time period prescribed by the sausage oscillation. Standing and propagating sausage oscillations are found to have different signatures in ZP patterns. We conclude that ZP wiggles can be used for the detection of short-period sausage oscillations and the exploitation of their seismological potential.
\end{abstract}


\section{Introduction}\label{sec-1}

Magnetohydrodynamic (MHD) waves in the solar corona are subject to intensive observational and theoretical research, especially in the context of diagnostics of coronal plasmas and {the} processes operating there \citep[e.g.][for recent comprehensive reviews]{2012RSPTA.370.3193D, 2014SoPh..289.3233L,2016SSRv..tmp....2N}. Interpretation of observed coronal oscillatory phenomena is usually based upon the MHD modes of a wave-guiding plasma non-uniformity. One of the most popular models of such a non-uniformity is a plasma cylinder stretched along the magnetic field \citep[e.g.][]{zaj75, 1983SoPh...88..179E}, which is often used for {studying} MHD oscillations in coronal loops and plumes, prominence fibrils, etc. 
Properties of different MHD modes are determined by {the} parameters of the waveguide, and also by the type of the mode. In a plasma cylinder, the axisymmetric fast magnetoacoustic mode is known as the sausage or peristaltic or $m=0$ mode, where $m$ is the azimuthal wave number. 
{This} mode is essentially compressive, and produces axisymmetric perturbations of the plasma density and frozen-in magnetic field. In the low-$\beta$ plasma {that is} typical for the corona, the sausage mode is characterised by alternate radial f{lows of the plasma, perpendicular to the field}.

Recent theoretical {studies} of the sausage mode include the study of the dependence of its parameters, e.g. the period and damping time, on the parameters of the oscillating loop, e.g. the density contrast, {the transverse profile of the fast magnetoacoustic speed} \citep[see, e.g.][]{2007A&A...461..1149P, 2012ApJ...761..134N, 2014A&A...567A..24H, 2014ApJ...781...92V, 2014A&A...572A..60L,2015ApJ...801...23L, 2015ApJ...814...60Y}; accounting for the effects of multi-shell structuring \citep[e.g.][]{2007SoPh..246..165P, 2015SoPh..290.2231C}, a field-aligned plasma flow \citep[e.g.][]{2014SoPh..289.1663C, 2014A&A...568A..31L}, non-uniform cross-section \citep{2009A&A...494.1119P}, and magnetic field twist \citep[e.g.][]{2014AnGeo..32.1189B, 2014SoPh..289.4069M}. Parameters of standing sausage modes provide us with a unique information about the transverse profiles of the plasma in flaring loops and the {external} Alfv\'en speed\citep[e.g.][]{2007AstL...33..706K, 2008PhyU...51.1123Z, 2009A&A...503..569I, 2015ApJ...812...22C}. {These quantities are} crucial for answering the enigmatic question of solar coronal heating \citep[e.g.][for a recent review]{2015RSPTA.37340269D}. Special attention {has been} paid to forward modelling of {the} characteristic features {of sausage oscillations, which appear in observations} in different bands, including imaging observations of the optically thin and thick thermal and non-thermal emission \citep{2012A&A...543A..12G, 2013A&A...555A..74A, 2014ApJ...785...86R, 2015A&A...575A..47R, 2015SoPh..290.1173K}.

One of the difficulties in the observational detection of the sausage mode is its short period. In typical flaring loops sausage oscillations have periods shorter than a few tens of seconds, which is shorter than the time resolution of EUV imagers traditionally used for coronal MHD seismology. But, if excited in the impulsive phase of a solar flare, this mode can be detected in the microwave and hard X-ray bands with ground-based and space-borne instruments that {have} high time resolution and sufficient spatial resolution \citep[e.g.][]{2003A&A...412L...7N, 2008A&A...487.1147I, 2010SoPh..267..329K, 2015A&A...574A..53K}. Also, it is believed that the sausage mode can be responsible for some quasi-periodic pulsations in solar and stellar flaring light curves \citep[see][for a review]{2009SSRv..149..119N, 2010PPCF...52l4009N}.
Recently, \cite{2013A&A...550A...1K} developed a model interpreting radio fiber bursts in terms of propagating sausage wave trains. 

Recently, signatures of standing sausage modes were found in moderate polarized zebra-patterns (ZPs) detected in flaring radio emission \citep{2013ApJ...777..159Y}. ZP are sets of almost locally parallel stripes in the dynamical spectra of broadband type IV radio bursts \citep[see][for a review]{2006SSRv..127..195C}. According to the most commonly accepted interpretation of this phenomenon, ZPs are produced by the coherent generation of upper-hybrid waves at multiple double plasma resonance (DPR) layers in a non-uniform plasma penetrated by non-thermal electrons accelerated in a solar flare \citep{1975SoPh...44..461Z, 2007SoPh..241..127K}. Different stripes are generated at different spatial locations \citep{2011ApJ...736...64C} where the DPR conditions are fulfilled, and the frequency separation of adjacent stripes is proportional to the magnetic field strength \citep[e.g.][]{2015A&A...581A.115K}. Frequencies of individual stripes correspond to the local electron plasma frequency, or its second harmonic, in the emission region. Hence, a magnetoacoustic wave, in particular the sausage mode,  in a flaring loop filled in with non-thermal electrons, could affect the corresponding ZP by periodic variation of the electron cyclotron frequency and plasma frequency. {This effect will produce} wiggles in the dynamical spectrum of the flaring radio emission. The periods of the ZP wiggles should correspond to the {periods} of the modulating magnetoacoustic waves. 

The aim of this work is to model the process of the ZP modulation by a sausage mode, providing a theoretical foundation for the interpretation of ZP wiggles in terms of MHD oscillations, and the  use of ZP wiggles for MHD seismology. In this study we concentrate on the MHD processes only, which determine the instant location of the sources of ZP emission stripes and their frequencies. The paper is organised as follows. Section~\ref{zpf} briefly describes the mechanism for ZP formation, based on the double plasma resonance.
The numerical MHD model is introduced in Section~\ref{MHDsims}. 
Results of numerical simulations are described in Section~\ref{results}.
Section~\ref{disc} presents the discussion and conclusions.

\section{ZP Formation}
\label{zpf}

According to the DPR model \citep[e.g.][]{1975SoPh...44..461Z, 2007SoPh..241..127K} the emission associated with individual ZP stripes comes from the spatial locations where {the following condition is fulfilled,}
\begin{equation}
\label{eqn:DPR}
s\omega_\mathrm{ce}=\sqrt{\omega_\mathrm{ce}^2+\omega_\mathrm{pe}^2}=\omega_\mathrm{uh},
\end{equation}
where $\omega_\mathrm{ce} = (e B/m_\mathrm{e})^{1/2}$ is the electron cyclotron frequency, $\omega_\mathrm{pe} = (n_\mathrm{e}e^2/m_\mathrm{e}\varepsilon_0)^{1/2}$ is the electron plasma frequency, $B$ is the absolute value of the magnetic field, $n_\mathrm{e}$ is the electron concentration in the background plasma, and $s$ is an integer representing the DPR harmonic number. In this model, {the radio emission is produced at the upper-hybrid frequency $\omega_\mathrm{uh}$, or its second harmonic, due to non-linear transformation of the plasma waves generated by} kinetic instabilities. {These instabilities are} caused by non-thermal electrons accelerated in the solar flare and streaming through the plasma. These instabilities are greatly enhanced under the DPR condition. In a plasma with non-uniform $B$ and $n_\mathrm{e}$, e.g. in a flaring loop or arcade,  the layers where condition (\ref{eqn:DPR}) is satisfied for different integers $s$, have different spatial locations \citep[see, e.g. Fig.~14 of][for illustration]{2011ApJ...736...64C}. Thus, different stripes of a ZP radio burst come from different regions of {the} flaring active region, depending on the specific value of the integer $s$. 

In the presence of a standing or propagating magnetoacoustic wave that perturbs the magnetic field $B$ and the plasma density, the spatial locations of the DPR layers and the resonant values of $\omega_\mathrm{uh}$ change in time. In this study, we determine the location and frequencies ($\omega_\mathrm{uh}$) of different DPR layers in different phases of standing and propagating sausage fast magnetoacoustic oscillations, taking into account non-uniformities of the plasma density and magnetic field both along and across the field.

\section{Magnetohydrodynamic numerical simulations}
\label{MHDsims}

\subsection{Equilibrium}
\label{equi}

\begin{figure}[htbp!] 
\epsscale{1.0}
\plotone{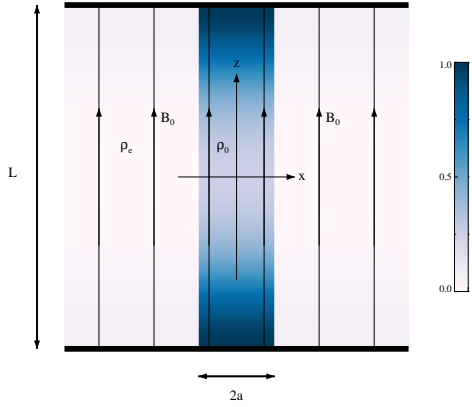}
\caption{Sketch of the equilibrium state. The colour shows the normalised local plasma density. The origin of the Cartesian coordinate set is at the centre of the computational domain. 
(A colour version of this figure is available in the online journal.) 
\label{fig-1}
}
\end{figure}

Following the standard approach introduced by \cite{1982SoPh...76..239E} we model a semi-circular coronal loop as a slab of  enhanced plasma density, directed along a uniform, constant magnetic field ${B}_\mathrm{0} = B_\mathrm{0}\hat{{z}}$ (see Figure~\ref{fig-1}). The local angle between
the magnetic field and the gravitational acceleration varies along the slab, allowing us to account for the loop's curvature \citep[see, e.g.][]{2000A&A...362.1151N}.
The profile of the plasma density in the transverse direction is given by a step-function of width $2a$. In the $z$-direction, along the field, the density is exponentially stratified and is described as a function of the height $h(z)=\frac{L}{\pi}\cos(\frac{\pi}{L}z)$,
\begin{equation}
\rho_0(x,z)=
\begin{cases}
	\rho_\mathrm{0}\exp(-{h(z)}/{H}) &    |x| < a, \\
	\rho_\mathrm{e}\exp(-{h(z)}/{H}) &    |x|  > a
\end{cases}
\end{equation}		
where $z$ is the coordinate along the magnetic field, with $z=0$ at the loop apex, $L$ is the loop length. For simplicity, we take the effective stratification to be the same inside and outside the loop. In the following we assume that the density scale height $H$ is comparable to the half length of the loop, $H= {L}/{2}$. This, a sufficiently small value of the scale height, corresponding to a low temperature in the case of a hydrostatic equilibrium along the field, is not crucial for the results obtained, and is needed only for better illustration of the results. 
Moreover, in our study, the instant density scale height does not need to coincide with the hydrostatic scale height that is prescribed by the plasma temperature, as the background plasma does not need to be in a hydrostatic equilibrium. As we are interested in short-term phenomena occurring in the impulsive phase of flares, with the typical time scale {of} about the fast wave travel time across the slab \citep{2014A&A...567A..24H}, the field-aligned flows that operate at a much {slower}, acoustic time scale, can be neglected. For the sausage mode, because of the existence of the ignorable coordinate, the slab and cylinder models give very similar results, as has been shown analytically by \citet{1982SoPh...76..239E} and \citet{1983SoPh...88..179E}, and numerically by \citet{2012ApJ...761..134N, 2014A&A...567A..24H}.

The magnetic field and plasma number density at $z=0$ were chosen as $25~\mathrm{G}$ and $4.6\times10^{10}~\mathrm{cm^{-3}}$, respectively. Thus the electron plasma and cyclotron frequencies are $1.9~\mathrm{GHz}$ and $70~\mathrm{MHz}$, respectively. These specific values are needed for the illustration purposes only, and can be readily changed by simple re-normalisation. The density contrast was chosen to be $\rho_\mathrm{0}/\rho_\mathrm{e}=10$. Thus the Alfv\'en speed inside and outside the loop, $C_\mathrm{A0}$ and $C_\mathrm{Ae}$ are $0.4\;\mathrm{Mm\,s^{-1}}$ and $1.1\;\mathrm{Mm\,s^{-1}}$, respectively, which is typical for solar coronal active regions. The plasma $\beta$ was taken to be $0.1$. This value is typical for flaring active regions, and it allows us to separate the scales of fast and slow magnetoacoustic waves. In the following study, we consider fast waves only. This approach is justified by {the} short duration of ZP, shorter than several seconds. During this time interval slow magnetoacoustic waves that have much longer periods (e.g. 3--10 minutes) do not produce any significant redistribution of the equilibrium density and magnetic field.

\subsection{Governing equations}

Numerical simulations of magnetoacoustic oscillations perturbing the equilibrium plasma configuration described in Section~\ref{equi} were performed using the \textit{Lare2D} code \citep{2001JCoPh.171..151A}, in terms of ideal MHD equations,
\begin{eqnarray}
\frac{\partial\rho}{\partial t} = -\nabla\cdot(\rho\mathbf{v}), \\
\frac{\mathrm{D}\mathbf{v}}{\mathrm{D} t} = \frac{1}{\rho}(\nabla\times \mathbf{B})\times \mathbf{B} - \frac{1}{\rho}\nabla P, \\
\frac{\mathrm{D}\mathbf{B}}{\mathrm{D}t} = (\mathbf{B}\cdot\nabla)\mathbf{v}-\mathbf{B}(\nabla\cdot\mathbf{v}), \\
\frac{\mathrm{D}\mathbf{\epsilon}}{\mathrm{D} t} = -\frac{P}{\rho}\nabla\cdot\mathbf{v}, \\
\nabla\cdot\mathbf{B}=0 
\end{eqnarray}
where the operator ${\mathrm{D}}/{\mathrm{D}t}$ denotes the total derivative, 
$\rho$ is the mass density, 
$\mathbf{v}$ is the velocity, 
$\mathbf{B}$ is the magnetic field, 
$\epsilon$ is the internal energy, 
$P=\rho\epsilon(\gamma -1)$ is the thermal pressure and $\gamma=5/3$ is the ratio of specific heats. 
{The} effect of the gravity on MHD oscillations was neglected. 

All the simulations were carried out in a domain of ($-L/2$, $L/2$)$\times$($-L/2$, $L/2$) (see Fig.~\ref{fig-1}), covered by $1000\times1000$ uniformly distributed grid points. The convergence was tested by increasing the grid density by a factor of 2.

\subsection{Boundary and initial conditions}

The equilibrium was perturbed by a pulse of the transverse component of the plasma velocity,  
\begin{equation}\label{equ-1}
v_x(x,z,t=0) = A_0\, x\,\exp\left[-\left(\frac{x}{\lambda_x}\right)^2 \right]\,f(z)
\end{equation}
where $A_0$ {and} $\lambda_x$ are the pulse amplitude and transverse width, respectively; and the function $f(z)$ describes the pulse structure in the longitudinal direction, along the field. The pulse is of the sausage symmetry, as the transverse component of the velocity is an odd function, which is zero at the axis of the slab.
In all simulations we {used a pulse width that coincides} with the slab half-width, $\lambda_x=a$. {This} guaranteed the preferential excitation of the {fundamental sausage mode in the transverse direction} \citep[see, e.g.][]{2004MNRAS.349..705N, 2005SSRv..121..115N, 2009A&A...503..569I, 2015ApJ...814..135S}. 

The function $f(z)$ was taken to be even. Its specific {form} allowed us to {excite} either propagating or standing sausage modes. Standing waves were excited by setting 
\begin{equation}\label{equ-2}
f(z) = \cos (kz),
\end{equation}
where $k = \pi/L$. In all simulations the pulse amplitude $A_0$ was chosen to be $0.1$, minimising non-linear effects that are not of interest in this study. This choice of $f(z)$ led to the preferential excitation of the {fundamental sausage mode in} the longitudinal direction. Propagating waves were excited by setting
\begin{equation}\label{equ1.1}
f(z) = \exp\left[-\left({z}/{\lambda_z}\right)^2 \right],
\end{equation}
where $\lambda_z \ll L$. Thus, in this case the initial pulse was localised in the vicinity of $z=0$, near the loop top, and developed towards the footpoints.

In all simulations performed in this study the length of the loop $L$ was chosen to be 10~Mm. The boundary conditions were line-tying in the magnetic field direction at $z= - L/2,~L/2$, 
corresponding to the photospheric boundary conditions at the coronal base. 
In the transverse direction, open boundary conditions were applied. 

\subsection{Simulations of MHD oscillations}
\label{results}

\begin{figure*}[htbp!]
\epsscale{2.0}
\plotone{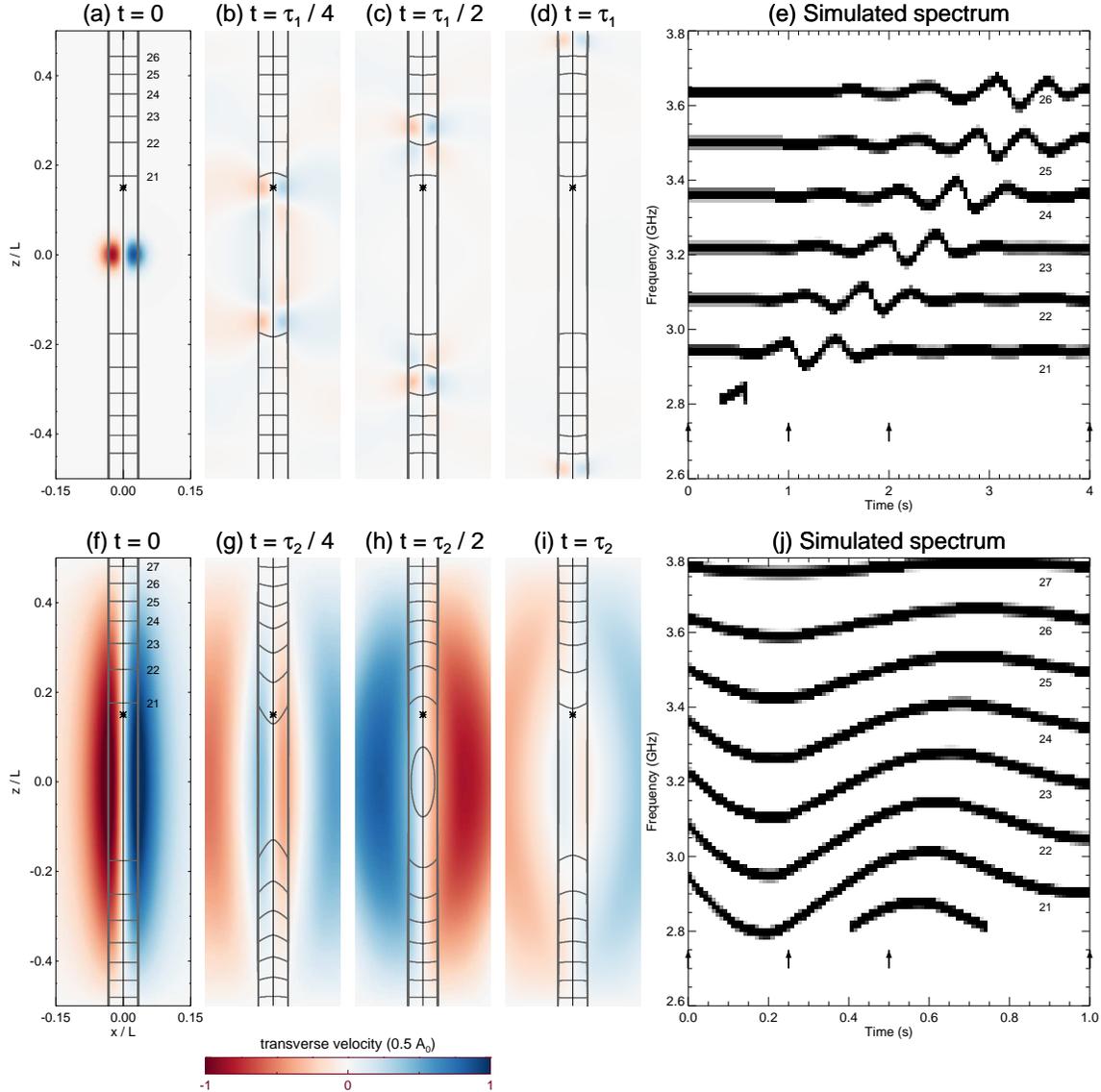}
\caption{ 
Evolution of fast magnetoacoustic pulses in a field-aligned plasma slab in the case of a propagating wave (panels a, b, c and d)  and a standing wave (panels f, g, h and i) and its possible manifestation in radio bursts as zebra-pattern wiggling. The colour shows the instant local values of the $x$-component of velocity, with the blue and red colours corresponding to the motions in the positive and negative directions, respectively. 
The snapshots are taken at times of $0, {1}/{4}, {1}/{2}, 1$ of the simulation runtimes, showing different phases of the oscillations. The simulation runtimes $\tau_1$ and $\tau_2$ were 4~s and 1~s for the propagating and standing waves, respectively. The grey contours in the snapshots show the position of the DPR levels. Panels e and j show the simulated dynamical spectra of the radio emission containing ZPs. The vertical arrows indicate the four time steps at which the snapshots are taken. The {asterisks} in panels (a--d) and (f--i) indicate the locations where the time-dependent signals shown in Fig.~3 are taken. The cyclotron harmonic numbers of the DPR levels and the corresponding ZP stripes are shown in panels a, e, f and j.
(A colour version of this figure is available in the online journal.) 
\label{fig-2}}
\end{figure*}

\begin{figure}[htbp!]
\epsscale{1.0}
\plotone{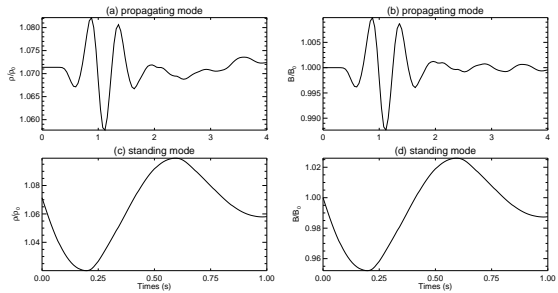}
\caption{Time profiles of the normalised density and magnetic field strength of the fast-propagating mode (first row) and the fast-standing mode (second row) at the point marked with the asterisk in Fig.~\ref{fig-2}.
\label{fig-3}}
\end{figure}

Figure~\ref{fig-2} shows the development of initial perturbations, visualised by snapshots of the spatial structure of the transverse flows. The upper row presents the evolution of the perturbation that is initially localised in the longitudinal direction at the centre of the slab (\ref{equ1.1}), along the slab axis. Panels \textit{a}, \textit{b}, \textit{c} and \textit{d} show different instants of time of the evolution, until the perturbation reaches the footpoints. The perturbation splits into two pulses, propagating in opposite directions along the slab in {the} form of fast magnetoacoustic waves. It is evident that a significant part of the energy is guided by the slab, as it {undergoes} a decrease in the fast speed. We observe the gradual development of an oscillatory wave train, with several cycles of oscillation along the slab, similar to that detected in numerical experiments of \citep[see, e.g.][]{2004MNRAS.349..705N, 2005SSRv..121..115N, 2010ITPS...38.2243J, 2014ApJ...788...44M, 2015ApJ...814..135S}. This phenomenon is caused by the geometrical dispersion caused by the plasma non-uniformity \citep{1984ApJ...279..857R}. In Figure~\ref{fig-3} we show time profiles of the developed wave trains. The perturbations of the plasma density and absolute value of the magnetic field are seen to be in phase, which corresponds to a fast magnetoacoustic wave in the low-$\beta$ plasma considered here. 

The bottom row (panels \textit{f}, \textit{g}, \textit{h} and \textit{i}) of Fig.~\ref{fig-2} {shows} different phases of {a} standing sausage oscillation, given by initial condition (\ref{equ-2}). As in the previous case, the oscillation consists of alternate converging and diverging transverse flows {to and from} the axis of the slab, and is accompanied {by} periodic perturbations of the density and magnetic field. The velocity oscillations have nodes at the footpoints ($z/L=\pm 0.5$)  and a maximum near the centre of the slab ($z/L=0$). This geometry corresponds to the fundamental (or global) sausage mode of the slab. Fig.~\ref{fig-3} shows time profiles of the oscillation taken at a specific point inside the loop. The plasma density and absolute value of the magnetic field oscillate almost in phase.

\subsection{Variations of the DPR layers}

As both propagating and standing sausage oscillations of the slab perturb the density and magnetic field, one could expect that {at} different instants of time, corresponding to different phases of the oscillations, the spatial locations of the DPR layers given by Eq.~(\ref{eqn:DPR}) are different. {This} is based upon the fact that in the low-$\beta$ regime the density and magnetic field strength are perturbed almost in phase, while the upper hybrid frequency depends on the magnetic field strength, via the electron cyclotron frequency, and the root of the plasma density, via the electron plasma frequency (see Eq.~(\ref{eqn:DPR})). Fig.~\ref{fig-2} shows variations of the locations of DPR in the loop caused by the sausage perturbations, at different instants of time corresponding to different phases of the oscillations or development of the running fast magnetoacoustic pulse. 
It is seen that the DPR frequencies corresponding to different harmonics $s$ vary in time. Thus, fast waves perturb the locations and values of different {stripes} in the generated ZP. 

The effect of the sausage oscillation modulation of ZP is demonstrated in the simulated spectra of ZP shown in the right panels of Fig.~\ref{fig-2} for propagating and standing waves. 
Skeletons of the radio emission spectra caused by the DPR mechanism were calculated with the use of Eq.~(\ref{eqn:DPR}). 
{We assume that the emission is produced at the loop axis only, i.e., the non-thermal electrons are concentrated within a very narrow slab around the loop axis. Then these electrons interact with the time-varying non-uniform plasma, and produce the radio burst in the intersections of the loop axis with the DPR layers at the double harmonics of the DPR frequencies.} 
We also assume that the emission intensity is only sensitive to the emission frequency, as the actual emission intensity depends on the processes that are beyond the MHD modeling (see Sec.~\ref{disc}). For visualisation purposes, a Gaussian line shape is adopted for different DPR emission lines corresponding to the individual ZP stripes. The standard deviation of the Gaussian function is set to be 30 MHz. It is several times larger than typical spectral resolution of, e.g. the  Solar Broadband Radio Spectrometer (SBRS) \citep[see, e.g.][]{1993SoPh..147..203J, 2004SoPh..222..167F} that is often used for ZP detection.
It is also evident that the simulated ZP spectra of the propagating and standing fast waves show propagating or standing patterns, respectively, in the radio spectrum.

\section{Discussion and Conclusions}
\label{disc}

We performed numerical modelling of the evolution of plasma parameters in a coronal loop by an axisymmetric, sausage fast magnetoacoustic oscillation, and determined the evolution of the DPR layers in the loop. Assuming that radio ZP structures are produced by the DPR mechanism, our study demonstrates that wiggles of ZP structures in broadband radio bursts can be produced by the variation of the macroscopic plasma parameters by a magnetoacoustic wave. We found that the modulations of ZP spectra, produced by propagating and standing sausage modes are different. In the case of the modulation by a standing wave the ZP stripes wiggle in phase, while for a propagating wave, e.g. formed due to the geometrical dispersion, the propagating pattern is clearly seen in the ZP spectrum. 
In particular, our modelling managed to reproduce the quasi-periodic wiggling of ZP stripes, {which was} observationally found by \cite{2013ApJ...777..159Y}. The period of the ZP wiggling is prescribed by the standing sausage mode period that, in turn, is determined by the parameters of the oscillating plasma non-uniformity \citep[e.g.][]{2012ApJ...761..134N}. Thus, the analysis of ZP wiggling allows one to get information about short-period MHD oscillations of the source region. These oscillations can be used for seismological diagnostics of the plasma in flaring regions \citep[e.g.][]{2012RSPTA.370.3193D,2014SoPh..289.1663C, 2015SoPh..290.2231C}.

This study has several important shortcomings. In particular, in our modelling we considered only a step-function transverse profile of the plasma density in the coronal loop. In this geometry, DPR layers are almost perpendicular to the local magnetic field, and different layers are situated along the field. This simplifying assumption was necessary because it allowed us to highlight the basic effect responsible for the ZP modulation. However, in the case of smooth profiles of the equilibrium plasma parameters (density, temperature, magnetic field) across the field, different DPR layers are situated not only across the field, but also along it, e.g. at the transverse slopes of loops and plumes, etc. Consideration of the effect of the smooth transverse structuring on the ZP modulation is an interesting future task. The MHD part of the study could also be advanced by including effects of the loop's curvature and cylindrical geometry, {for example}. 

Furthermore, in this study, we restricted our attention to the MHD modelling only, determining the \lq\lq skeleton\rq\rq\ of the ZPs and their modulation, based on the DPR condition. Our approach was similar to the modelling performed by \cite{2013A&A...550A...1K} for radio fiber bursts. In {the present} study, this approach was chosen too, as we aimed to theoretically reproduce the results obtained by \citet{2013ApJ...777..159Y} who studied the variation of the ZP stripe skeletons only.
The emission of electromagnetic waves by the interaction of non-thermal electrons with the plasma was not considered because it requires one to account for non-MHD effects, which would  complicate the modelling and possibly hide the important physics involved. The possible time and spatial variation of the non-thermal electron distribution function \citep[e.g.][]{2013SoPh..284..579Z} was neglected too. Likewise, we did not consider the possible  evolution of the ZP emission polarisation \citep[e.g.][]{2014SoPh..289..233Z, 2015ApJ...808L..45K}. Accounting for these effects may provide us with additional seismological information, and hence should be addressed in future studies.


\acknowledgments
We thank the anonymous referee for his/her comments and suggestions which helped to improve this paper. The work is supported by NSFC grant No. 11221063 and 11273030, MOST grant No. 2011CB811401, and the National Major Scientific Equipment R\&D Project ZDYZ2009-3. This research was also supported by the Marie Curie PIRSES-GA-2011-295272 \textit{RadioSun} project, the European Research Council under the \textit{SeismoSun} Research Project No. 321141, STFC consolidated grant ST/L000733/1, and the BK21 plus program through the National Research Foundation funded by the Ministry of Education of Korea (VMN).

\bibliographystyle{apj}
\bibliography{ms}

\end{document}